\documentclass[twocolumn,superscriptaddress,longbibliography,aps,pra,preprintnumbers]{revtex4-1}

\usepackage[normalem]{ulem}
\usepackage{graphicx}
\usepackage{bm}
\usepackage{color}
\usepackage{epstopdf}
\usepackage{amsmath}
\usepackage{amssymb}
\usepackage{epstopdf}
\usepackage{lipsum}

\usepackage[urlcolor=blue,colorlinks=true,citecolor=blue,linkcolor=blue,pdfstartview={FitH},bookmarks=false]{hyperref}

\graphicspath{{fig/}{./fig/}{.}}

\begin{document}

\title{
Electronic properties of Bi$_{2}$Se$_{3}$ \\
dopped by $3d$ transition metal (Mn, Fe, Co, or Ni) ions
}

\author{Andrzej Ptok}
\email[e-mail: ]{aptok@mmj.pl}
\affiliation{Institute of Nuclear Physics, Polish Academy of Sciences, ulica W. E. Radzikowskiego 152, PL-31342 Krak\'{o}w, Poland}

\author{Konrad Jerzy Kapcia}
\email[e-mail: ]{konrad.kapcia@ifj.edu.pl}
\affiliation{Institute of Nuclear Physics, Polish Academy of Sciences, ulica W. E. Radzikowskiego 152, PL-31342 Krak\'{o}w, Poland}

\author{Anna Ciechan}
\email[e-mail: ]{ciechan@ifpan.edu.pl}
\affiliation{Institute of Physics, Polish Academy of Sciences, Aleja Lotnik\'{o}w 32/46, PL-02668 Warsaw, Poland}

\date{\today}

\begin{abstract}
Topological insulators are characterized by the existence of band inversion and the possibility of the realization of surface states.  
Doping with a magnetic atom, which is a source of the time-reversal symmetry breaking, can lead to realization of novel magneto--electronic properties of the system.
In this paper, we study effects of substitution by the transition metal ions (Mn, Fe, Co and Ni) into Bi$_{2}$Se$_{3}$ on its electric properties.
Using the {\it ab inito} supercell technique, we investigate the density of states and the projected band structure.
Under such substitution the shift of the Fermi level is observed.
We find  the existence of nearly dispersionless bands around the Fermi level associated with substituted atoms, especially, in the case of the Co and Ni.
Additionally, we discuss the modification of the electron localization function as well as charge and spin redistribution in the system.
Our study shows a strong influence of the transition metal--Se bond on local modifications of the physical properties.
The results are also discussed in the context of the interplay between energy levels of the magnetic impurities and topological surface states.
\end{abstract}

\maketitle

\section{Introduction}
\label{sec.intro}

The spin-orbit interaction leads to the realization of a topological insulator (TI)~\cite{hasan.kase.10,qi.zhang.11}.
Similarly as in the case of ordinary insulators, TIs have a band gap, which separates the highest occupied electronic band from the lowest empty band.
However, the surface of a TI is characterized by gapless states protected by time-reversal symmetry~\cite{kane.mele.05,kane.mele.05b,moore.balents.07,roy.09,roy.09b}.
This phenomena was experimentally observed for the first time in HgTe / CdTe quantum wells~\cite{zhang.liu.09,nechaev.hatch.13}.

Bi$_{2}$Se$_{3}$ is one of several large band-gap three dimensional (3D) TIs~\cite{zhang.liu.09}.
In this compound a topologically non-trivial direct gap of $0.3$~eV is observed~\cite{nechaev.hatch.13}.
These topological properties were predicted by {\it ab initio} calculation~\cite{xia.qian.09} and are associated with band inversion near $\Gamma$ point~\cite{noh.koh.08}.
As a consequence, the realization of the Dirac cone surface states with hexagonal deformation~\cite{alpichshev.analytis.10} is observed~\cite{xia.qian.09,hsieh.qian.08,hsieh.xia.09b,hsieh.xia.09,hsieh.xia.09c,kuroda.arita.10,nomura.souma.14}.

Electronic parameters of Bi$_{2}$Se$_{3}$ compound can be  modified in a relatively simple way.
For instance, the surface states in the slab geometry can be tuned by strain~\cite{aramberri.munoz.18}.
Moreover chemical gating is used to shift the Fermi level to the the spin-degenerate Dirac point~\cite{hsieh.xia.09b} or above the conduction band bottom edge in the case of higher concentration of doped ions~\cite{hor.richardella.09}.
Similar behavior is observed also for the magnetic doping~\cite{chen.chu.10}.
Finally, doping can modify the spin--orbit coupling, as observations via angle-resolved photoemission (ARPES) measurments show~\cite{king.hatch.11}.

\begin{figure}[!b]
\centering
\includegraphics[width=0.7\linewidth]{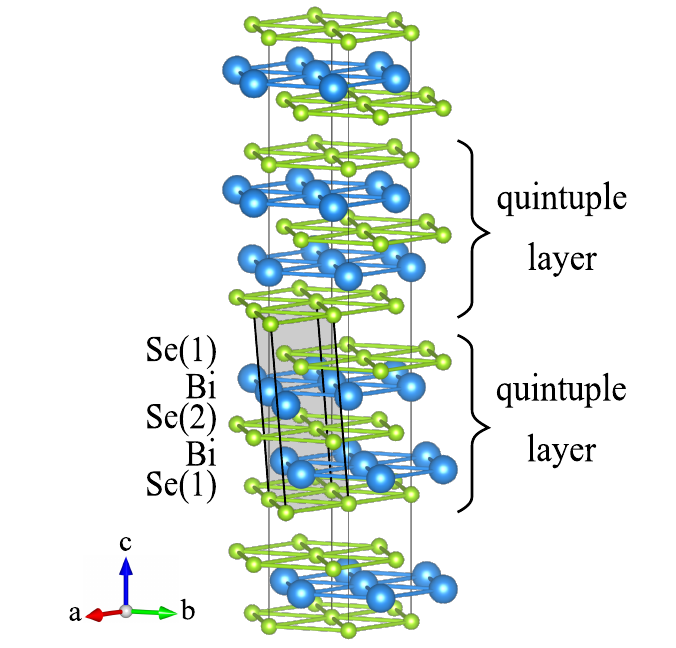}
\caption{
Crystal structure of Bi$_{2}$Se$_{3}$ topological insulator.
A primitive cell including one formula unit is marked by gray shadow and thick black lines, whereas a conventional unit cell (containing three primitive cells) is denoted by thin black lines.
The image was rendered using {\sc Vesta} software~\cite{momma.izumi.11}.
\label{fig.schemat}
}
\end{figure}

In the case of the magnetic adatom on the surface of Bi$_{2}$Se$_{3}$, a giant magnetic anisotropy is observed (e.g., for Co~\cite{gambardella.rusponi.03,eelbo.sikora.13} or Fe~\cite{honolka.khajetoorians.12,eelbo.wasniowska.14}).
Additionally, such an impurity breaks the time-reversal symmetry and suppresses the low-energy local density of states (LDOS) around the impurity.
The surface states mediate a coupling between the impurities, which is always ferromagnetic, whereas the chemical potential lies close to the Dirac point.
Therefore, we expect that a finite concentration of magnetic impurities should give rise to a ferromagnetic ground state on the surface~\cite{liu.liu.09}.
This mechanism provides the physical realization of the novel topological magneto-electric effect.

Studies of the TI and magnetic insulator (MI) heterostructures~\cite{eremeev.menshov.13,luo.qi.13} suggest a charge redistribution and mixing of the TI and MI orbitals at the interface, what leads to a drastic modification of the electronic structure near the TI/MI interface.
This suggests the existence of the interfacial ordinary state confined within the adjacent interfacial quintuple layer of the TI, which slowly decays into the MI. 
This state is shifted downwards to the local energy gap owing to the near-interface band bending. 
Moreover, this state is gaped and spin-polarized due to the hybridization with the MI states.

The impact of the substituted atom is also observed in the scanning tunneling microscope (STM) surface measurements~\cite{dai.west.16,miao.xu.18}.
Even if the impurity is located below the surface, the charge redistribution in the form of an electronic defect can be observed~\cite{mann.west.13}.

Ideal Bi$_{2}$(Se,Te)$_{3}$ reveals bulk insulating behavior, 
however, crystals grown from the stoichiometric melts possess rather a metallic character~\cite{jia.ji.11}.
This is a consequence of the presence of native defects like Bi antisites or Se vacancies which shift the Fermi level into the conduction band (they are created due to the small defect formation energies)~\cite{ren.taskin.12}.
In this situation, the downshift of the Fermi level to the topological gap region can be realized by doping of the system with nonmagnetic atoms like Ca, Mg or Sn~\cite{hsieh.qian.08,kuroda.arita.10,kushwaha.gibson.14}.
Moreover, for relatively large doping, the Fermi level can be shifted to the conduction band.

In this paper, we discuss the role of doping of topological insulator Bi$_{2}$Se$_{3}$ on its properties. 
We study the effects caused by the $3d$ transition metal (TM = Mn, Co, Fe, and Ni) ions substituted at the Bi site.
We investigate several quantities such as the density of states (DOS), band structures, electron localization functions (ELF), and magnetic and charge distributions.
We show that doping by TM atoms leads to the existence of nearly dispersionless band associated with the substituted atom, which in the case of Co and Ni is located around the Fermi level.
In contrast to the nonmagnetic atoms, the magnetic TM breaks the time reversal symmetry.
As a result, novel topological properties of the system are expected~\cite{sanchezbarriga.varykhalov.16}.
Finally, we discuss this aspect in the context of the possible experimental arrangement.

\begin{table}[!b]
\caption{
\label{tab.dft}
The calculated lattice constants (after optimization using different pseudopotentials) compared with experimental data. 
Results are obtained in the presence of the spin--orbit coupling and van der Waals correction.}
\begin{ruledtabular}
\begin{tabular}{lccc}
& Exp.~\cite{perez.vicente.99} & LDA & GGA \\
\hline
$a$ (\AA) & 4.1355 & 4.1996 & 4.1173 \\
$c$ (\AA) & 28.615 & 28.5180 & 27.7146 \\
$E_{g}$ (eV) & 0.3 & 0.285 & 0.270 \\
\end{tabular}
\end{ruledtabular}
\end{table}

The paper is organized as follows.
Sec.~\ref{sec.num_res} is devoted to the presentation of numerical results.
We briefly describe computational details (Sec.~\ref{sec.calc_det}),  present the results obtained for  clean (Sec.~\ref{sec.bi2se3_clean}) and  doped (Sec.~\ref{sec.bi2se3_doped}) systems.
Next, in Sec.~\ref{sec.conseq}, we discuss interplay between the doping and the surface states in the context of current experimental results.
Finally, a summary is included in Sec.~\ref{sec.sum}.

\begin{figure}[!t]
\centering
\includegraphics[width=\linewidth]{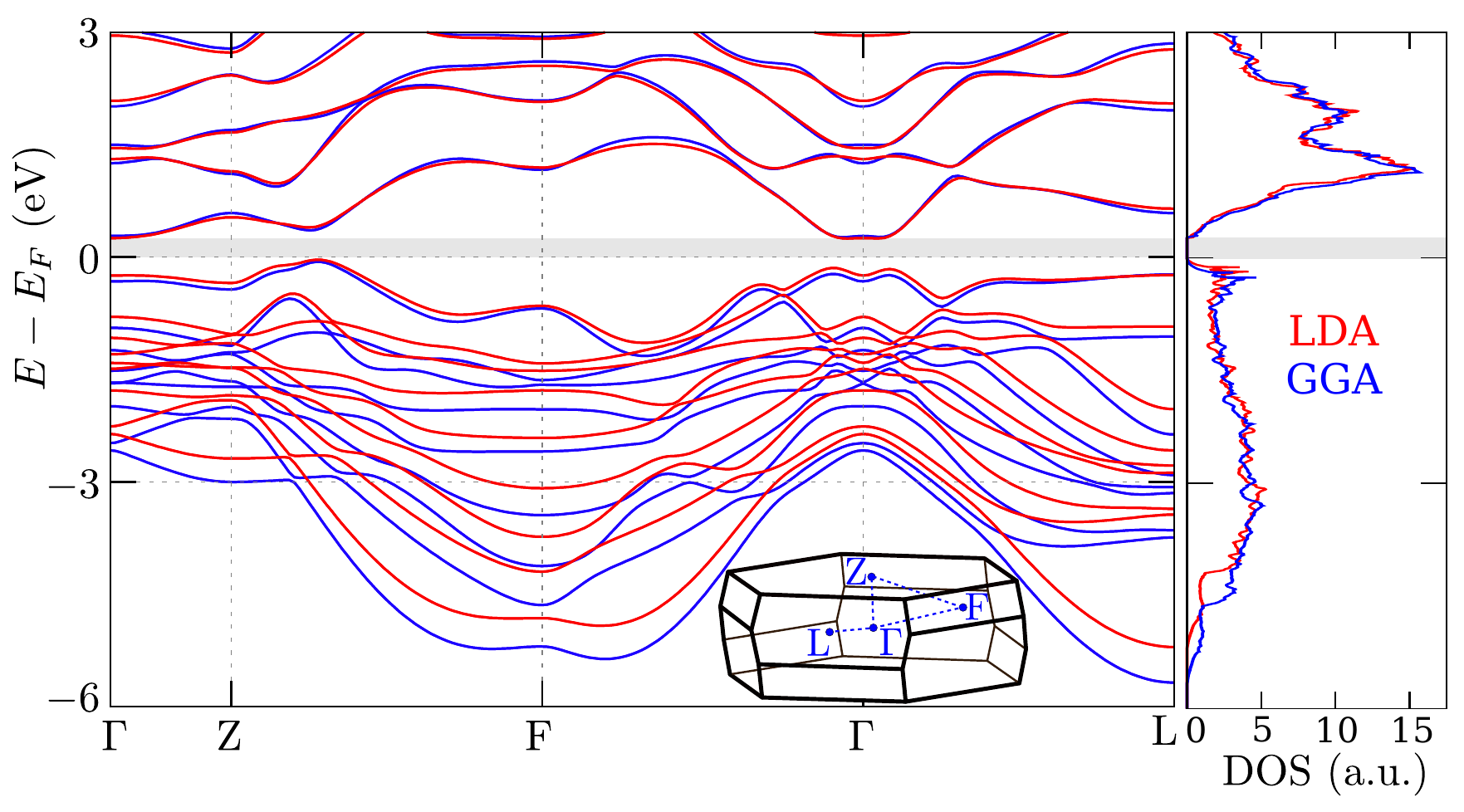}
\caption{
Band structure and density of states of clean Bi$_{2}$Se$_{3}$ in the case of LDA and GGA calculations (red and blue lines, respectively).
The topological band gap is marked by the grey background.
Cf. also Ref. ~\cite{zhang.liu.09}.
The results are obtained for the lattice constants after optimization (which are different for both cases, see Table~\ref{tab.dft}; calculation are performed in the primitive cell).
The Fermi level is located at zero energy.
\label{fig.clean}
}
\end{figure}

\begin{table}[!b]
\caption{
\label{tab.dft_supercell}
The lattice parameters (of $3 \times 3 \times 1$ supercell) for the system with different substitutions.
Magnetic moment (in $\mu_{B}$) denotes the value of magnetic moment localized on the TM atom.
}
\begin{ruledtabular}
\begin{tabular}{lcccccc}
& {\it clean} & Mn & Fe & Co & Ni \\
\hline
$a$ (\AA) & $12.599$ & $12.572$ & $12.572$ & $12.568$ & $12.571$ \\
$c$ (\AA) & $28.518$ & $28.484$ & $28.491$ & $28.458$ & $28.453$ \\
mag. mom. ($\mu_{B}$) & $0.000$ & $4.208$ & $3.641$ & $2.349$ & $0.443$ \\
\end{tabular}
\end{ruledtabular}
\end{table}

\begin{figure*}[!t]
\centering
\includegraphics[width=\linewidth]{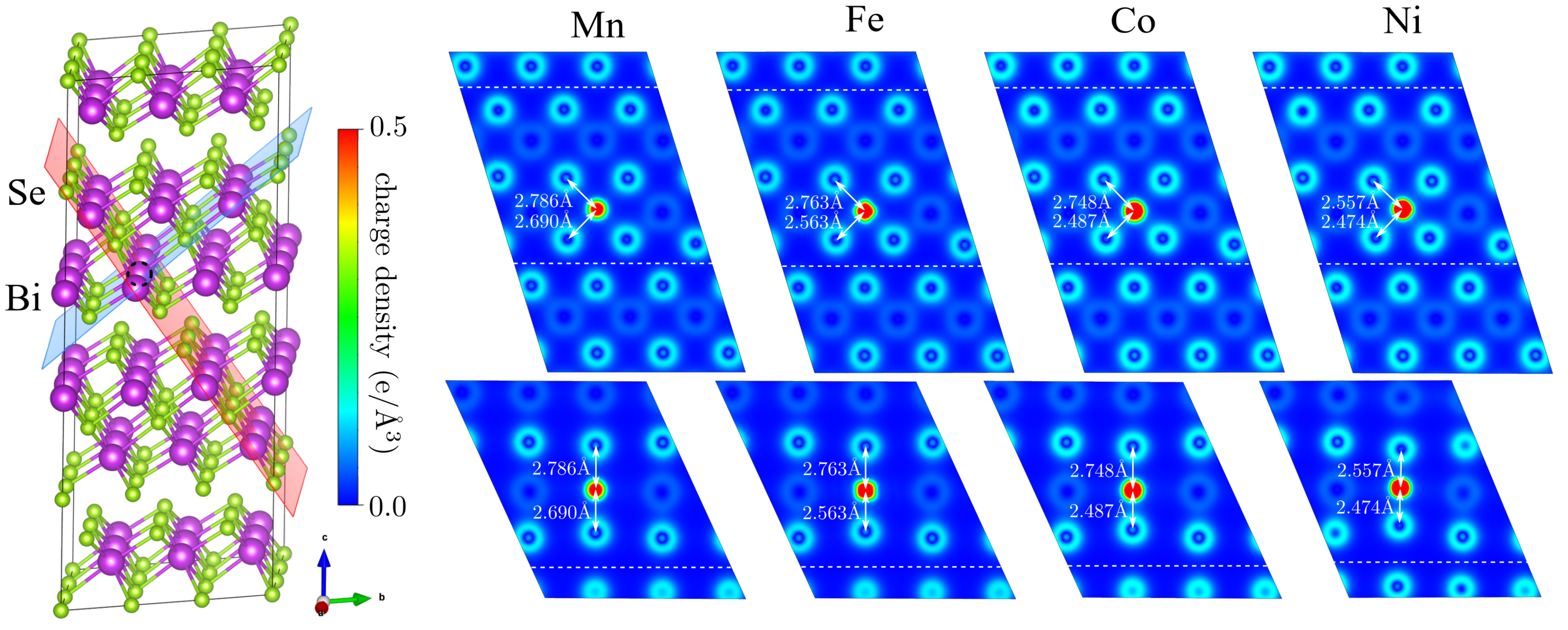}
\caption{
The electronic density in two different planes containing the substituted TM atom [marked by red and blue planes on the left panel showing $3 \times 3 \times 1$ supercell; upper (lower) panels correspond to the red (blue) plane from the left panel].
The TM atom is labeled on the top. 
Distances between the TM atoms and the nearest Se atoms are indicated, whereas the dashed white lines show a location of the van der Waals gap between the quintuple layers.
\label{fig.bound}
}
\end{figure*}

\begin{figure*}[!p]
\centering
\includegraphics[width=0.96\linewidth]{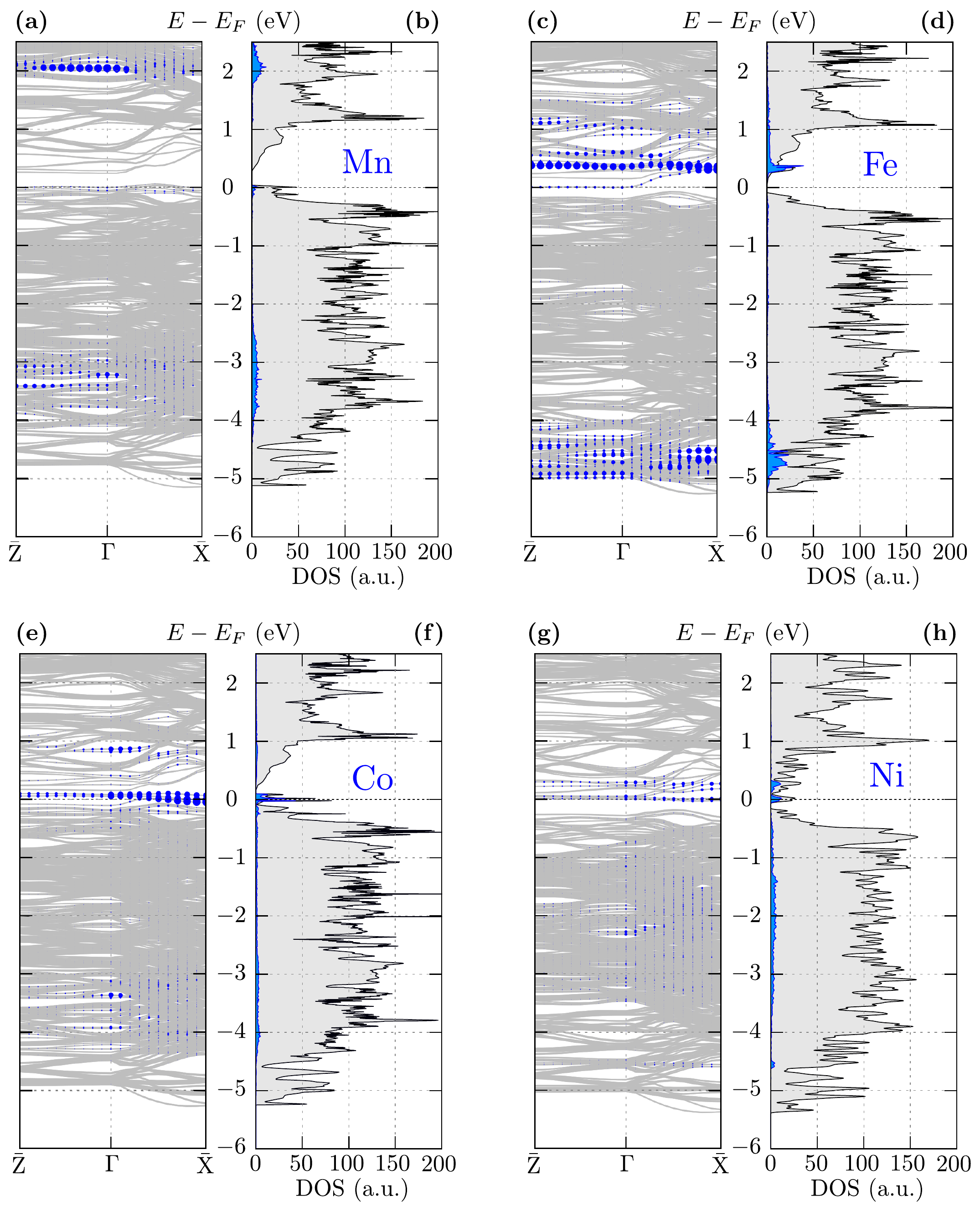}
\caption{
The band structures and the electronic density of states (DOS) for the TM doped systems (labeled on the top).
Band structure is plotted along high symmetry points in the supercell geometry [i.e., $\bar{\text{Z}} = (0,0,0.5)$, $\Gamma = (0,0,0)$, and $\bar{\text{X}} = (0.5,0,0)$].
In plots of the band structure [i.e., panels (a), (c), (e), and (g)], size of blue dots corresponds to the contribution of the substituted TM atom to the energy level.
Similarly,  blue lines in DOS [i.e., panels (b), (d), (f), and (i)] show partial DOSs of the substituted TM atoms.
The Fermi level is located at zero energy.}
\label{fig.dos}
\end{figure*}

\section{Numerical results and discussion}
\label{sec.num_res}

\subsection{Calculation details}
\label{sec.calc_det}

The first-principles calculations are preformed using the projector augmented-wave (PAW) potentials~\cite{blochl.94} implemented in the Vienna Ab initio Simulation Package ({\sc vasp}) code~\cite{kresse.hafner.94,kresse.furthmuller.96,kresse.joubert.99}.
The calculations are made within the local density aproximation (LDA)~\cite{perdew.zunger.81,perdew.wang.92} and the generalized gradient approximation (GGA) in the Perdew, Burke, and Ernzerhof (PBE) parametrization~\cite{pardew.burke.96}.
First, the crystal structure is optimized in the conventional unit cell including three primitive cells (cf. Fig.~\ref{fig.schemat}, where the primitive cell is marked by gray shadow) with the spin--orbit coupling (SOC).
As a break of the optimization loop, we take the condition with an energy difference of $10^{-4}$~eV and $10^{-6}$~eV for ionic and electronic degrees of freedom.
During the crystal structure optimization loop, ionic positions, cell volume, and cell shape are allowed to relax~\footnote{
Analogous calculations can be performed in the case of the lattice parameters and atomic position obtained from the experimental data.
However, the absence of the volume relaxation leads to the emergence of internal pressure. 
In our case we found: $-12.82$~kB for the LDA and $29.00$~kB for the GGA.
Similarly, the absence of the atomic relaxation leads to the emergence of internal forces acting on atoms. 
The maximal force found in the case LDA is equal to $0.2215$~eV/\AA, while for GGA it is $0.2986$~eV/\AA.
Also the gap value is underestimated and it is equal to approximately $0.25$~eV in both cases.
Moreover, in the case of the doped system, the absence of the optimization step can lead to unphysical results and behaviors due to incorrect distances between atoms.}.
Van der Waals (vdW) corrections are included using the Grimme scheme (DFT-D2)~\cite{grimme.06}.
In the case of the doped system, the $3 \times 3 \times 1$ supercell (containing $27$ primitive cells) is used with one TM impurity substituting Bi site. 
It corresponds to the concentration of magnetic impurities of about $1.85$\%.
For the summation over the reciprocal space, one uses $12 \times 12 \times 3$ and $4 \times 4 \times 3$ {\bf k}--point $\Gamma$--centered grids in the Monkhorst--Pack scheme~\cite{monkhorst.pack.76}, in the case of the conventional cell (for the clean system) and the supercell (for the doped system), respectively.
The energy cutoff for the plane-wave expansion is set to $350$~eV.
The crystall symmetry is analyzed using {\sc FindSym}~\cite{stokes.hatch.05} and {\sc SpgLib}~\cite{togo.tanaka.18}, while the momentum space analysis is done with using {\sc SeeK-path} tools~\cite{hinuma.pizzi.17}.

\subsection{Clean Bi$_{2}$Se$_{3}$}
\label{sec.bi2se3_clean}

The Bi$_{2}$Se$_{3}$ family of compounds crystallizes with the $R\bar{3}m$ symmetry (space group no. $166$)~\cite{perez.vicente.99}.
As shown in Fig.~\ref{fig.schemat}, the system has a layered structure with the quintuple layer (QL) as a basic unit.
Each QL consists of five atomic layers formed by Se(1)--Bi--Se(2)--Bi--Se(1) atoms.
Within the QL, Se atoms take two nonequivalent positions.
Se(1) atomic layer is adjacent only to one Bi layer.
Contrary to this, Se(2) atomic layer located inside the QL is surrounded by two Bi layers.
Inside the QL, strong bonding between molecular $p$-like orbitals  is produced along the five-atom linear chain~\cite{mishra.satpathy.97}.
The interaction between two Se(1) layers belonging to two different QLs layers is of the van der Waals type.

\begin{figure*}[!t]
\centering
\includegraphics[width=\linewidth]{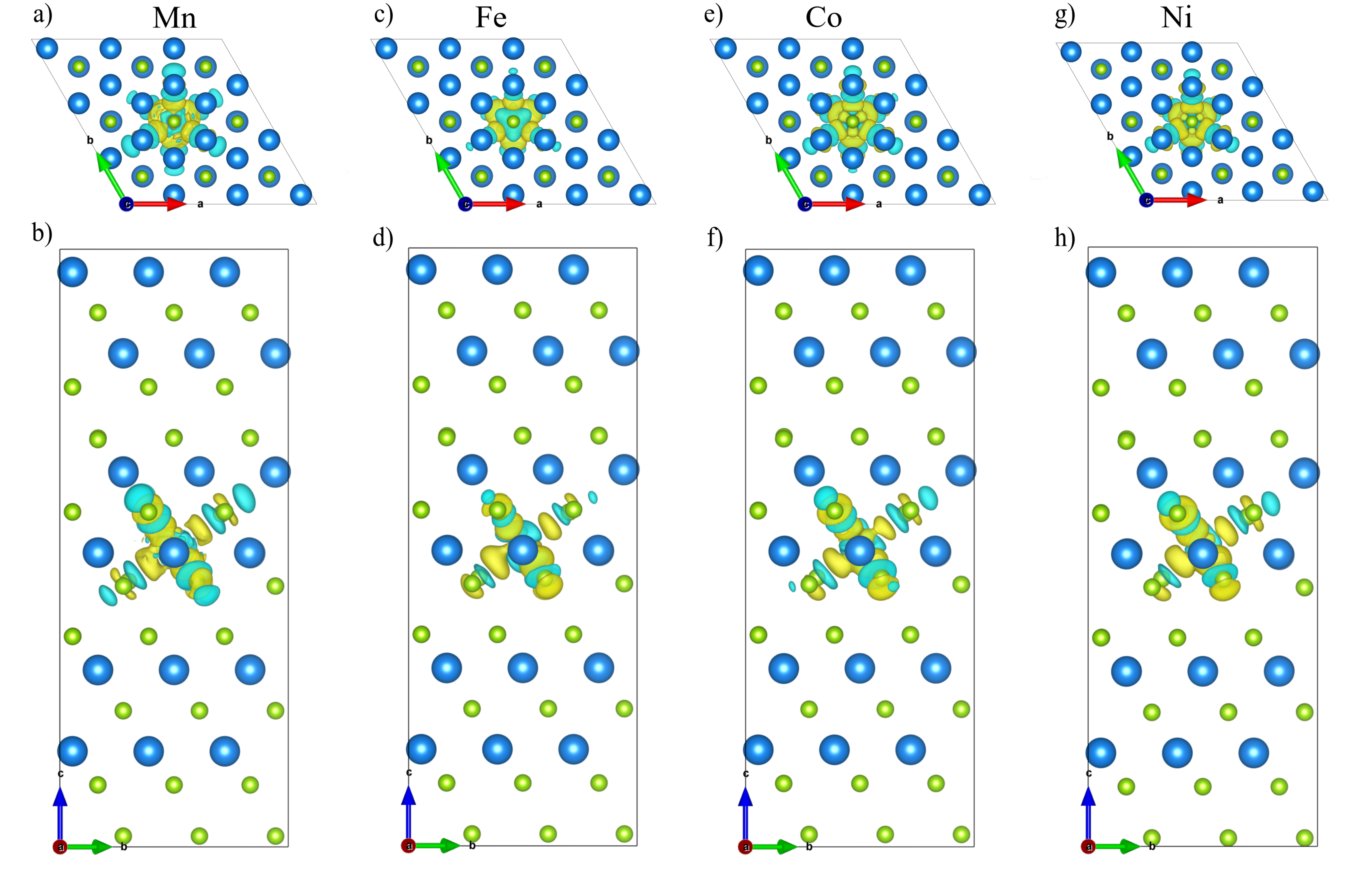}
\caption{
Modifications of the charge distribution by the substituted TM atoms (for isosurfaces absolute modification equaling $2 \times 10^{-3}$~e/\AA$^{3}$).
Yellow and cyan colors correspond to increasing and decreasing charge density, respectively. 
The TM atom is labeled on the top.
\label{fig.chg}
}
\end{figure*}

In the case of Bi$_{2}$Se$_{3}$, lattice parameters and gap values are very sensitive to used pseudopotentials~\cite{park.ryu.16}.
For instance, some hybrid pseudopotentials lead to underestimation of gap values, while GGA PBE gives an incorrect indirect gap~\cite{crowley.tahirkheli.15}.
In this compound, interactions of van der Waals type play a crucial role~\cite{luo.sullivan.12,shirali.shelton.19}.
Previous studies showed, that the fullest consistency of experimental and theoretical results can be obtained from LDA+$GW$ calculations with the relaxed atomic positions~\cite{nechaev.hatch.13,aguilera.friedrich.13,michiardi.aguilera.14}.
In addition, theoretical studies show existence of the strong correlation between the value of the topological gap and the value of lattice constant $c$~\cite{aramberri.munoz.18}.

A comparison of our results in the case of LDA and GGA PBE pseudopotentials is shown in Table~\ref{tab.dft} and Fig~\ref{fig.clean}. 
In both cases, band structures and DOS have qualitatively similar form.
However, the LDA calculations reproduce both the experimental parameters and the band gap of the clean Bi$_{2}$Se$_{3}$ more accurately (i.e., without substitutions).
Therefore, in the following parts of the manuscript, we present the results obtained solely by using the LDA pseudopotentials.

\subsection{Doped Bi$_{2}$Se$_{3}$}
\label{sec.bi2se3_doped}

In TM-doped Bi$_{2}$Se$_{3}$, the optimization of the crystal structure leads to the lowering of the supercell symmetry to the $P3m1$ (space group no. $156$), due to the distortion of atoms located near the substituted TM atom.
Additionally, distance between the TM atom and the nearest Se atoms in the doped system decreases in the series  Mn, Fe, Co, Ni and, simultaneously, it is smaller than the distance between Bi--Se atoms in one QL of the clean system,
due to smaller atomic radius of considered TMs atoms than Bi.
We present explicitly values of these distances in Fig.~\ref{fig.bound}.
As we can see, the TM--Se(1) distance (to the van der Waals gap site) is more suppressed than the TM--Se(2) bond (directed to the QL center). 
In the clean system, these relaxed distances are equal $2.903$~\AA\ and $3.102$~\AA, respectively.
Experimental data for the clean system indicates $2.850$~\AA\ and $3.074$~\AA, respectively.
Moreover, the strong bonding between TM and Se atoms is visible in the charge density.
This shows a strong local contraction of the bonds, which has been also observed in similar TM-doped structures~\cite{figueroa.vanderlaan.15,ghasemi.kepaptsoglou.16,ruzicka.caha.15}.
This contraction enhances the covalent character of the TM--Se bonds, which can have important consequences for the electronic and magnetic properties of the TM-doped Bi$_{2}$Se$_{3}$~\cite{figueroa.vanderlaan.15}.

The results of our calculations show that the supercell volume of the doped system is smaller than that in the clean case (cf.~Table~\ref{tab.dft_supercell}).
One should note that this behavior was reported previously in the case of the Mg doped system~\cite{tokarz.zalecki.14}.
In the case of the Sb substitution, the volume initially increases for low doping and then decreases for high doping concentration~\cite{devidas.amaladass.17}.
If Bi$_{2}$Se$_{3}$ is doped by TMs, magnetic moment appears in the system which is localized mainly on TM ions. Due to hybridization with the rest of the states, we observe smaller magnetic momentsthan for the corresponding free atoms.
The fractional magnetic moments are discussed in Sec.~\ref{sec.conseq} in more detail.

\paragraph*{Band structure and density of states.}---
Calculated band structures (BS) and density of states (DOS) are shown in Fig.~\ref{fig.dos}.
Projection of states into the TM substituted atoms are marked by blue dots and blue background area, in the case of BS and DOS, respectively.
In every case, we observe many well-localized dispersion-less states of the TM atoms.
However, only for  Co and Ni substitutions, we observe their localization around the Fermi level, e.g., in the form of clear peak at $E=0$~eV in DOS.
Typically, the system becomes an electron-doped and the Fermi level is shifted to the bottom of the conduction band~\cite{wei.zhang.15}.
Similar behavior was observed experimentally in the case of MnBi$_{2}$Se$_{4}$/Bi$_{2}$Se$_{3}$ heterostructure~\cite{hirahara.eremeev.17}, i.e., a magnetic monolayer deposited into the TI.
A shift of Fermi level is associated with a change of the energy levels around $\sim 0.5$~eV~\cite{zhang.velev.16}, which is bigger than the related value of insulating gap.
Contrary to this, in the case of Mn-doped system, the Fermi level is located below the top of the valence band of the clean system.

The band structures reveals differences between the TM and the QL atoms.
First, energy levels of the bands of the base system (gray lines in Fig.~\ref{fig.dos}) exhibit weak momentum dependence in the $z$-direction  due to QL structure.
An impact of the TM is indicated by a size of the blue dots in Fig.~\ref{fig.dos}.
In each case, we notice  a ``wide'' energy window of the TM states below the Fermi level, what is related to strong hybridization of the $d$-orbital  energy levels (of TM) with $p$-orbital states.
As opposed to this, the TM states above or near the Fermi level are well-localized in ``narrow'' energy window in a form of dispersionless bands.
Generally, the shift of the energy between levels below and above the Fermi level is relatively concordant with the splitting due to the field associated with the magnetic moment of the TM atom (cf.~Table~\ref{tab.dft_supercell}).
From this, independently of the spin-orbit coupling, in the studied system, one expects the separation of electrons in spin-$\uparrow$ and spin-$\downarrow$ subspaces in the energy domain.

Similarly as in the clean system, the valence and conduction bands consist of Bi and Se $p$ orbitals~\cite{mishra.satpathy.97}.
All other orbitals are located far away from the Fermi level (in the energy domain). 
In the clean system, these $p$ orbitals are responsible for a creation of strong $\sigma$-bonding between Bi--Se atoms~\cite{mishra.satpathy.97}.
Introduced  substituted TM atom modifies the bonding in the frame of one QL, what is very well visible in modifications of the charge distribution (Fig.~\ref{fig.chg}).

\paragraph*{Charge and spin redistribution.}---
The changes in the charge distribution induced by the substituted TM atom can be specified more precisely using the differences of charge and spin densities.
The modification of the charge density in Bi$_{2}$Se$_{3}$ by substitution of TM atom, can be shown by the difference between the charge densities of the (initial) doped system and both separated components~\cite{sikora.kalt.19}, namely:
\begin{eqnarray}
\label{eq.charge_dens} \Delta \rho = \rho_\text{TM$_{x}$Bi$_{2-x}$Se$_{3}$} - \left( \rho_\text{TM$_{x}$/vac} + \rho_\text{vac/Bi$_{2-x}$Se$_{3}$} \right) ,
\end{eqnarray}
where 
$\rho_\text{TM$_{x}$Bi$_{2-x}$Se$_{3}$}$, $\rho_\text{TM$_{x}$/vac}$, and $\rho_\text{vac/Bi$_{2-x}$Se$_{3}$}$ correspond to charge density calculated for the doped system, separated TM atom, and the system without substituted atoms (i.e., the system in the presence of vacancy~\footnote{Here, the {\it system with one vacancy} should be understood as the supercell after relaxation without the substituted TM atom.}), respectively.
The redistribution of the spin density can be defined analogously. 
Modifications of the charge distribution due to the TM substitution are shown in Fig.~\ref{fig.chg}.
In both cases strongest modification of charge and spin (not shown here) densities are observed along strong bonds between TM and Se atoms. 
Here, it should be noticed that Se(1) atoms play more important role  than Se(2) atoms in the frame of one QL.
The redistribution of charge and spin inherits of the C$_{3}$ symmetry of the system.
This behavior can be observed indirectly in the STM mapping of the Bi$_{2}$Se$_{3}$ surface~\cite{mann.west.13,miao.xu.18,dai.west.16}. 
In this case, when vacancy or impurity is included to the system, modification of the charge is strongly observed along $\sigma$ bonding.
As a consequence characteristic charge density defects with C$_{3}$ symmetry on the surface is observed~\cite{mann.west.13,dai.west.16,miao.xu.18}.

Finally, one notes that the magnetic moment of the system is strongly localized on the TM atom (cf. Table~\ref{tab.dft_supercell}).
Thus, only a small modification (in a range of $1$~m$\mu_B$/\AA$^{3}$) of the spin density around atoms near the TM impurity is observed.

\paragraph*{Electron localization function.}---
The influence of the TM on the character of the electronic states around the substituted atom can be studied by the electron localization function (ELF)~\cite{becke.edgecombe.90,savin.jepsen.92,silvi.savin.94}, which can take values from $0$ to $1$.
Large values of ELF (i.e., close to $1$) is associated with a high localization (small mobility) of electron at a given place. 
Small values (i.e., around $0$) correspond to  delocalized (itinerant) electrons.

\begin{figure}[!t]
\centering
\includegraphics[width=0.8\linewidth]{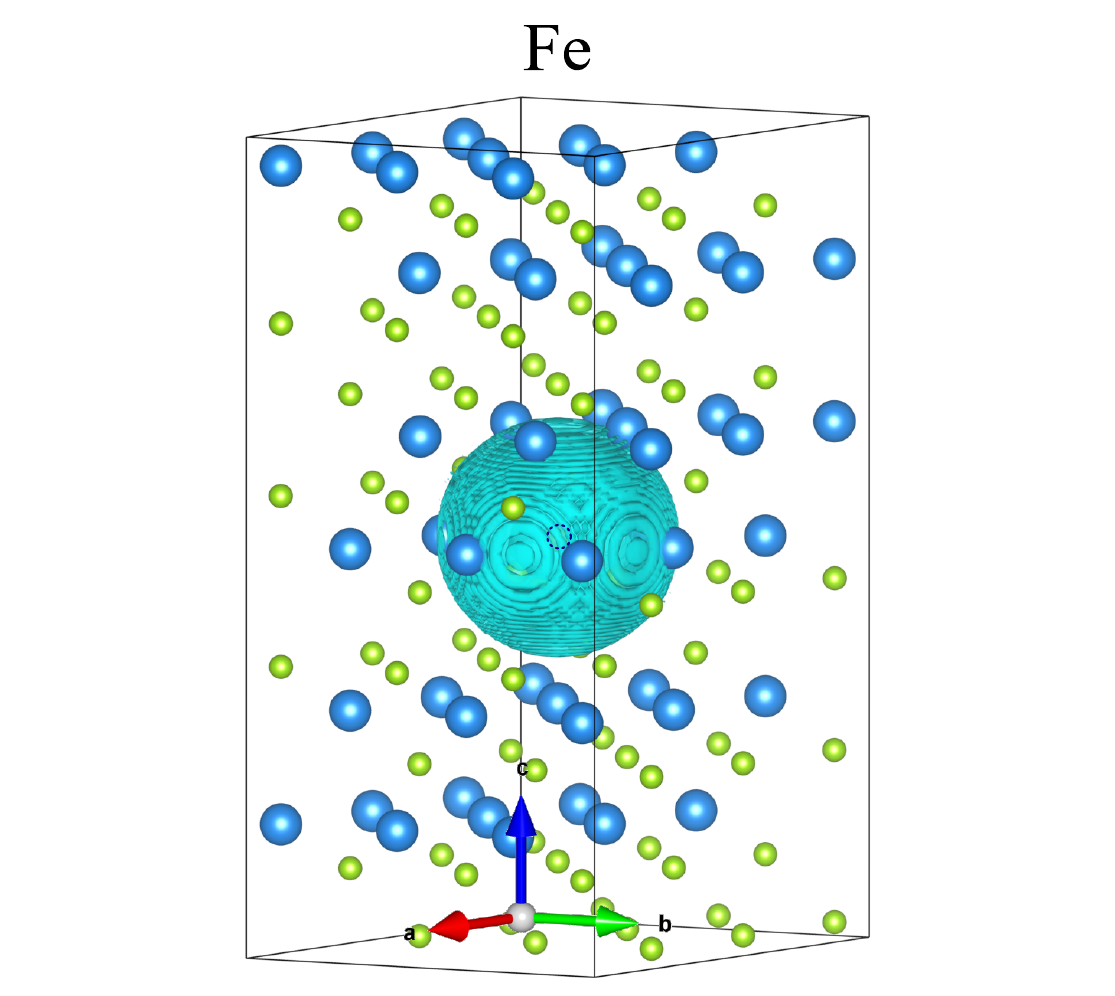}
\caption{
The modification of the electron localization function (ELF) by the Fe substitution [cf.~Eq.~(\ref{eq.delf})].
The TM atom is located at the center of the supercell.
Presented isosurface is obtained for $\Delta \text{ELF} = 0.5$.
\label{fig.elf}
}
\end{figure}

Similar as in the case of the charge density, the modifications of the ELF can be calculated as [cf. Eq.~(\ref{eq.charge_dens})]:
\begin{eqnarray}
\label{eq.delf} \Delta \text{ELF} &=& \text{ELF}_\text{TM$_{x}$Bi$_{2-x}$Se$_{3}$} \\
\nonumber &-& \left( \text{ELF}_\text{TM$_{x}$/vac} + \text{ELF}_\text{vac/Bi$_{2-x}$Se$_{3}$} \right) .
\end{eqnarray}
For instance, we show the ELF modification in the case of the Fe dopant (Fig.~\ref{fig.elf}).
The base system (i.e., clean Bi$_{2}$Se$_{3}$) is TI, which implies large value of  $\text{ELF}_\text{Bi$_{2}$Se$_{3}$}$ (nearly $1$ in whole volume of the system).
Introduction of the TM atom leads to decrease of the localization electrons around this atom. 
This behavior is compliant with experimental data, where the influence of magnetic impurity Fe on single crystals Bi$_{2}$Se$_{3}$ was investigated in the high magnetic fields by Shubnikov--de Haas (SdH) effect~\cite{sugama.hayashi.01}.
Based on the SdH effect and the Hall effect measurements, it was established that the doping of Bi$_{2}$Se$_{3}$ single crystals by Fe leads to an increase of the concentration of free electrons.

\subsection{Interplay between impurity levels and topological surface states}
\label{sec.conseq}

\begin{figure}[!b]
\centering
\includegraphics[width=\linewidth]{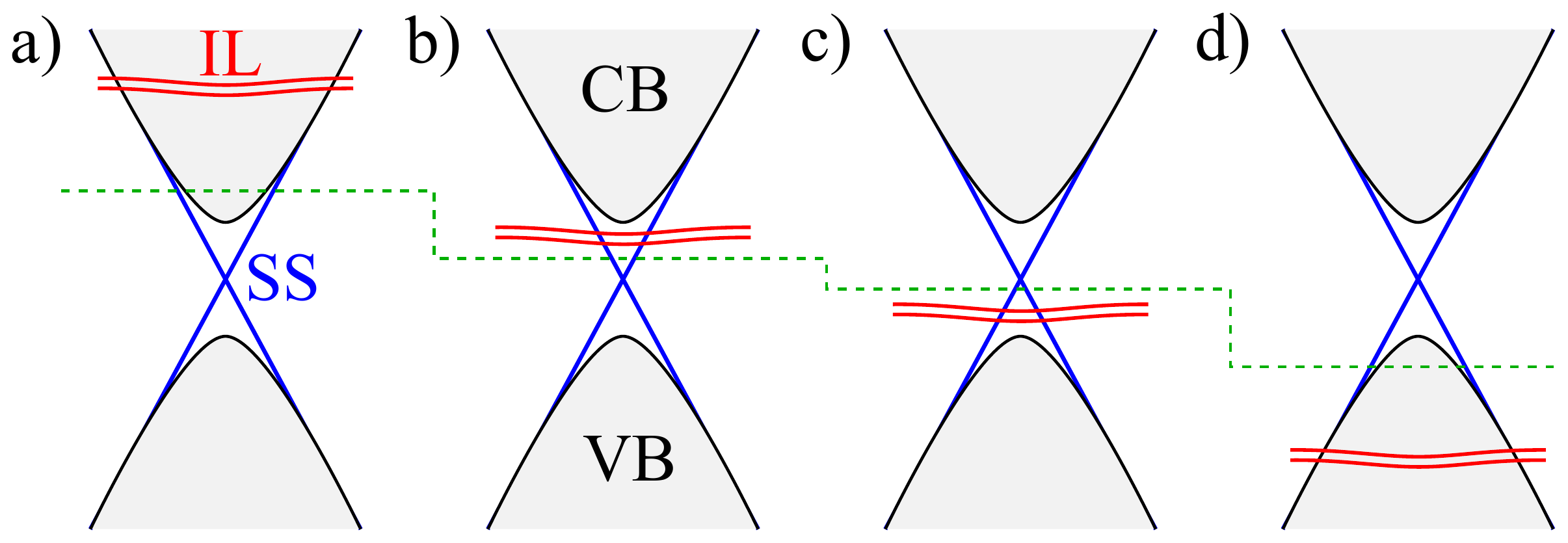}
\caption{
Schematic representation of the doped system band structure near the Fermi level (the green dashed line).
In the clean system one observes the topological gap between the continuum spectrum of the valence band (VB) and the conduction band (CB), represented by the black parabolas with the gray backgrounds.
Due to the topological properties of the TI, in a realistic system one observes forming of the metallic surface states in a form of the Dirac cones (the blue lines).
Introducing the impurity to the system leads to an emergence of the impurity levels (ILs), in a form of the (nearly-)dispersionless band (the red lines).
\label{fig.band_schem}
}
\end{figure}

In this subsection, we begin with discussing the impact of the impurity on the band structures in the general case.
The schematic representation of the band structure of the doped system is presented in Fig.~\ref{fig.band_schem}.
In the case of the clean bulk system, we observed existence of the topological gap (cf. Fig.~\ref{fig.clean}), between valence and conduction bands (represented by the black parabolas with the gray background in Fig.~\ref{fig.band_schem}).
Due to topological nature of the Bi$_{2}$Se$_{3}$, in a real system with the 'edge', one observes forming of the metallic surface states in the form of the Dirac cone (the blue lines in Fig.~\ref{fig.band_schem})~\cite{alpichshev.analytis.10,hsieh.qian.08,xia.qian.09,hsieh.xia.09,hsieh.xia.09b,hsieh.xia.09c,kuroda.arita.10,nomura.souma.14}.
Doping of the system leads to the introduction of the impurity levels (ILs) to the energy spectrum of the system (the red line in Fig.~\ref{fig.band_schem}).

Properties of the bands formed by the ILs depends on, e.g., the type of the impurity and its concentration.
The type of the impurity atoms plays an important role in the shifting of the Fermi level (schematically indicated by the green dashed line in Fig.~\ref{fig.band_schem}).
The single impurity leads to an appearance of discrete energy levels, i.e., flat bands, while increasing concentration of the impurities leads to forming of dispersive bands.
A theoretical study of the Bi$_2$Se$_3$ doped with Sn in the frame of the Korringa-Kohn-Rostoker method~\cite{wiendlocha.16}, shows that the change of the doped atom concentration from $0.05$~\% to $1$~\% leads to modification ILs from separate narrow peak to dispersion band with the width of about  $0.25$~eV.
A similar effect can be expected in the case of the TM atom doping.
In our case, increasing number of impurities can be associated with emergence of the sublattice of the TM atoms.
However, distances between them should be relatively large (of the order of several lattice constants).
In lower concentration, the mediation between dopant atoms should be carried out via itinerant electrons of the base system.
Increasing concentration should enhance second mechanism of dispersive band forming, i.e., increasing overlap between orbitals of the TM atoms within the sublattice.
Additionally, above some critical concentration, a realization of the long-range ferromagnetic (FM) order can be expected due to the Ruderman-Kittel-Kasuya-Yosida mechanism~\cite{carva.balaz.20}.
This effect was theoretically investigated, e.g., in the case of the Bi$_2$Se$_3$ doped with Mn, for concentrations above $3.2$~\%.
However, experimental data from the Mn doped Bi$_2$Se$_3$ epitaxial layer, suggest existence of long-range FM order even for smaller concentration of Mn (around concentration of 1.0~\%)~\cite{bardekeben.cantin.13}.

In the case of the atoms with a high (low) electron concentration, the Fermi level can be shifted to the conduction (valence) band [cf. Fig.~\ref{fig.band_schem}(a) and (d)].
One can expect a situation, in which ILs are located around the topological gap above or below the Fermi level [Fig.~\ref{fig.band_schem}(b) and (c), respectively].

Here, we should mention a role of the hybridization of the doped atom and atomic levels of the base system.
Dependently on a position of the energy levels of the TM doped atom, we observe different behavior of its DOS contribution (cf. Fig.~\ref{fig.dos}).
States around the topological gap are narrow, but states located deep in the bands are more smeared.
This can be associated with the hybridization of the TM orbitals with the orbitals of base system.
Exact analyses of the partial electronic DOS~\cite{tse.yu.15} show that the valence band is  composed mostly of $p$ states of Se, while conduction band is composed of $p$ states of Se and Bi as well as $s$ states of Se.
The $d$ states of TM are hybridized with $p$ orbitals of the neighboring Se atoms, which suppress their localization. 
If the ILs lie around the topological gap, the strong peaks appear in DOSs and become more delocalized as the dopant concentration increases. 
The situation is similar to ordinary semiconductors~\cite{raebiger.lany.09,masoumi.nadimi.18,ciechan.boguslawski.18}, in which the TM impurity levels can form clear peaks only if they are weakly hybridized with the host states of the same symmetry.


Doping of the Bi$_{2}$Se$_{3}$ by the magnetic atoms, leads not only to the Fermi energy level shift but also to breaking of the time reversal symmetry.
Thus, the gap is opened at the Dirac point (so-called {\it Dirac gap})~\cite{chen.chu.10,sanchezbarriga.varykhalov.16}.
Moreover, the spin texture of the surface states is modified~\cite{xu.neupane.12}.
This effects have consequences within the surface electric transport measurements~\cite{hsieh.xia.09b}, and will be discussed in the next paragraphs.

\paragraph*{The Fermi level in real system.}---
As we mentioned in the introduction, the Bi$_{2}$Se$_{3}$ crystal exhibits metallic character due to intrinsic defects~\cite{jia.ji.11,ren.taskin.12}, which have their donor levels above the conduction band minimum as in Fig.~\ref{fig.band_schem}(a).
This is in contrast to the theoretical results, where the TI nature of this compound is well visible (cf.~Sec.~\ref{sec.bi2se3_clean} and Fig.~\ref{fig.clean}).
In the simplest case, the shift of the Fermi level to the topological gap region can be obtained by doping of the system by nonmagnetic atoms.
For example, in the case of the 0.25\% doping by Ca atoms, the Fermi level was shifted to the Dirac point~\cite{hsieh.qian.08,hsieh.xia.09b}.
Similar effect was observed in the case of 1\% Mg doping~\cite{kuroda.arita.10}.
Further increase of the doped atom concentration leads to the shift of Fermi level to the conduction band~\cite{hor.richardella.09}.
Also some more complicated crystal structures are obtained, e.g., (Bi$_{1-x}$Sb$_{x}$)$_{2}$(Se$_{1-y}$Te$_{y}$)$_{3}$~\cite{arakane.sato.12,kushwaha.gibson.14} and other related doped compounds~\cite{kushwaha.platikosi.16,cheng.wu.16} to compensate all donors in the system and to lower the Fermi level.

A shift of the Fermi level to lower energies can be also carried out by doping the system by magnetic atoms. 
This shift strongly depends on a type of doped atoms.
For example, doping by Fe opens the Dirac gap, while the Fermi level is located in the conduction band. 
Replacing Fe by Mn atoms leads to a shift of the Fermi level to the lower energies~\cite{chen.chu.10,sanchezbarriga.varykhalov.16}.
ARPES experiments for the Mn doped system show the location of the Fermi level above the Dirac gap (more precisely, the Fermi level crosses the surface states from the conduction band) for Mn concentration smaller than~$0.5$~\%~\cite{sanchezbarriga.varykhalov.16}.
Then, the Mn energy levels are visible in the DOS at the top of the valence band~\cite{sanchezbarriga.varykhalov.16}.
This type of electron structure modification is the most interesting when it leads to the location of the Fermi level around the Dirac point and consequently to the realization of the quantum topological transport.
Experimental data suggests that this can be realized for Mn concentration of~$1.0$~\%~\cite{chen.chu.10}.

\begin{figure}[!t]
\centering
\includegraphics[width=\linewidth]{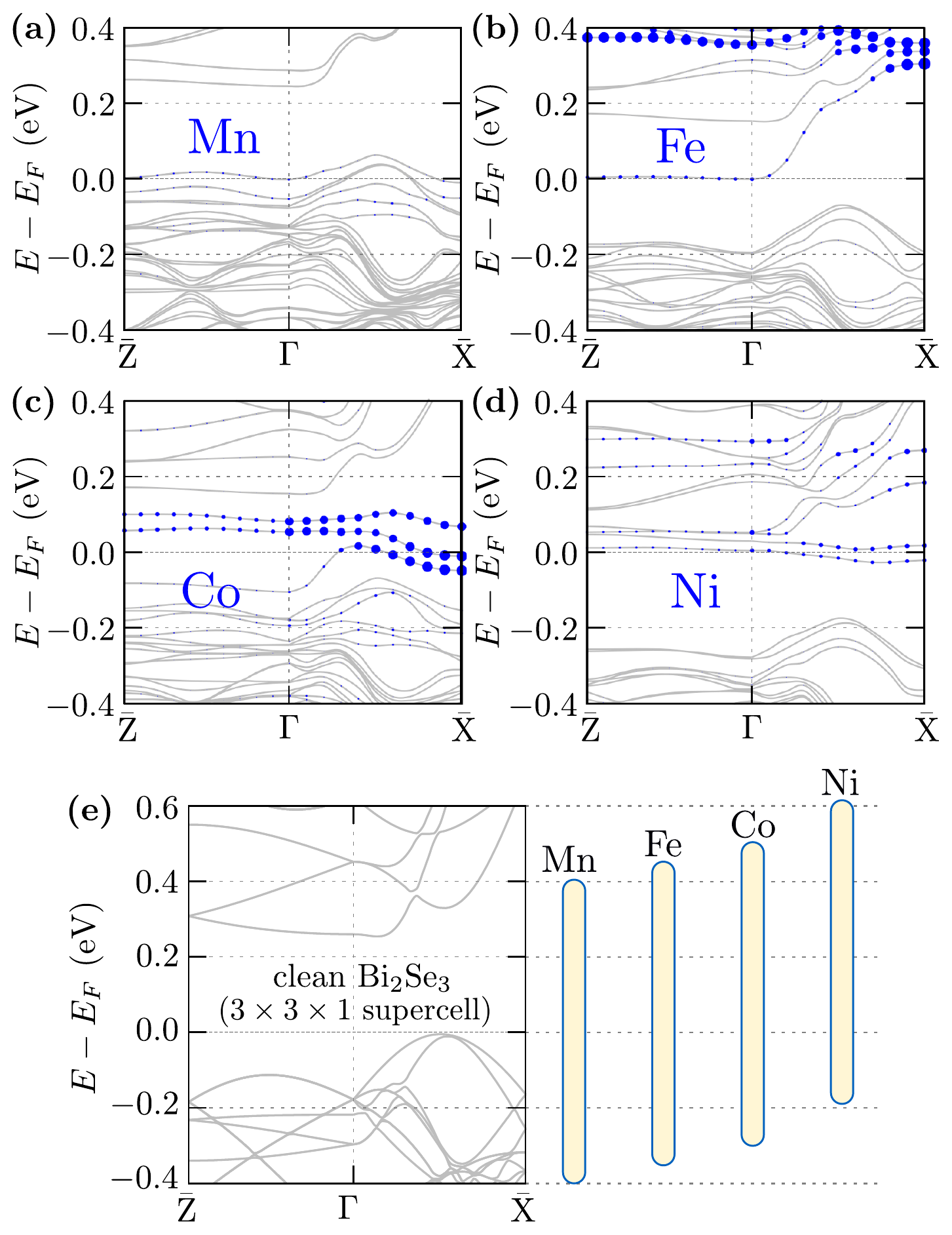}
\caption{
(a)--(d) Zoom on the electronic band structures around the Fermi level presented in Fig.~\ref{fig.dos} with a focus on a contribution of the substituted TM atom to energy levels.
All denotations are as in Fig.~\ref{fig.dos}
(i.e., size of blue dots corresponds to a contribution of the substituted TM atom to the energy level).
(e) The reference band structure of the clean Bi$_{2}$Se$_{3}$ calculated in the $3\times3\times1$ supercell.
Yellow bars on the right of panel (e) denote the energy ranges shown on panels (a)--(d).
\label{fig.band_small}
}
\end{figure}

The above-mentioned results are predominantly in the compliance with our {\it ab initio} theoretical results (see Fig.~\ref{fig.band_small} for details of the electronic band structures near the Fermi level). 
Doping of the system by Mn leads to the shift of the Fermi level to the valence band [Fig.~\ref{fig.band_small}(a)], while in all other cases the Fermi level is located in the conduction band.
Comparing the band structures obtained for the TM doped systems [panels (a)--(d) of Fig.~\ref{fig.band_small}] with the reference band structure of the clean Bi$_2$Se$_3$ [Fig.~\ref{fig.band_small}(e)], one notices that the band degeneracy is removed.
This is a consequence of two effects: {\it (i)} modification of the crystal structure by doped atoms (i.e., a change of the distances between atoms) and {\it (ii)} removal of the spin-degeneracy due to the time reversal symmetry breaking by magnetic TM dopants.
However, the location of the Fermi level depends also on the impurity atom concentration.

Here, we would like to discuss the role of the doping on the value of the  magnetic moment of the substituted TM atoms.
From the results collected in Tab.~\ref{tab.dft_supercell}, we can conclude that the magnetic moments are fractional and thus the calculated charge states of the TM ions are between $2+$ and $3+$ (except for Ni ions, in which the magnetic moment indicates the charge state between $3+$ and $4+$).
Taking into account that the real samples exhibit conductivity of $n$-type, we expect that they act as acceptors and exist in $2+$ charge state. 
Indeed, Mn$^{2+}$ signal with magnetic moment of $5 \mu_B$ in Bi$_2$Se$_3$ was observed by EPR measurements~\cite{wolos.drabinska.16}.
The Mn$^{2+}$ level is located within the valence band [see Fig.~\ref{fig.band_small}(a)] and thus compensates nonintentional donors and shift the Fermi energy below the conduction band.
Assuming that the main effect of increasing Mn concentration on the electron structure is broadening of the Mn-induced bands, the Fermi level should be shifted down to the Mn level (so to the  maximum of the valence band) as long as all donors are not compensated. 
For higher concentrations, both Mn$^{2+}$ (with maximal magnetic moment of $5$~$\mu_B$) and Mn$^{3+}$ (with $4$~$\mu_B$ magnetic moment) can coexist in the sample.

\paragraph*{Experimental observation.}---
An increase of the doped atom concentration can also lead to realization of the ferromagnetic order, which is induced by the spin-spin interaction mediated by the surface states~\cite{liu.liu.09,chen.chu.10}.
In fact, the magnetic impurities lead to the effect similar to the external magnetic field, where negative magnetoresistance (MR) is observed~\cite{wang.yan.15}.
In the absence of magnetic doping, the negative MR is due to the weak localization effect coexisting with the weak anti-localization effect (positive MR) under a low magnetic field~\cite{cheng.wu.16,lang.he.13,cha.classsen.12,yue.rule.20}.
The transition between weak localization and weak anti-localization is demonstrated as a gap opening at the Dirac point of surface states in the quantum diffusive regime.
This phenomena was reported in the case of the Cr doped TI~\cite{liu.zhang.12}.
This type of experiments, realized in the TI nanoflake devices, suggest the occurence of the surface dominated transport in low temperatures~\cite{xie.ren.13}.
Here, doping is an effective way to manipulate the magneto-transport properties of the TI~\cite{yue.rule.20}.
Also in the MR measurements, results strongly depend on the Fermi level location. 
For example, in the Fe doped material, the experiments show exactly a location of the Fermi level in the conduction band~\cite{sugama.hayashi.01}, what is in  concordance with results presented in this work.
Morover, in this case the negative MR was reported~\cite{singh.gangwar.19}.
In Bi$_{2}$Te$_{3}$, an increase of the Fe doping  leads to several transition on the phase diagram, from a paramagnetic TI, through a magnetic band insulator, to a valence bond glass TI~\cite{kim.kim.13}
Similar effect is expected in the Bi$_{2}$Se$_{3}$~\cite{singh.gangwar.19}.

Finally, an interesting phenomena associated with the topological properties of the magnetically doped 3D TI is the realization of the quantum anomalous Hall effect (QAHE)~\cite{yu.zhang.10,chang.zhang.13}.
When one spin channel is suppressed due to time reversal symmetry breaking, the other spin channel should exhibit QAHE in the form of the dissipationless spin chiral edge mode.
The time reversal symmetry can be broken due to magnetic doping, and thus the external magnetic field is not necessary~\cite{chang.li.16,fei.zhang.20}.
When the Fermi level is located around the Dirac point, the QAHE is expected.
Such conditions are provided by Mn$^{2+}$ or Fe$^{3+}$, see Fig.~\ref{fig.band_small}.
However, the presence of native donors makes it impossible to realize  Fe$^{3+}$ for any Fe concentration. 
Indeed, experimental realization of this phenomena was reported in the Mn~\cite{checkelsky.ye.12,liu.teng.18}, Cu~\cite{singh.kumar.20}, or Cr~\cite{pan.liu.20}, but not in Fe doped TI.

One should also notice that the mentioned  above experimental results were obtained in the case of the quasi-2D system, i.e., in a form of nanoflakes or nanodevices.
In such situations, the role of the magnetic doping is expected to be more important than in real 3D TI.

\section{Summary}
\label{sec.sum}

In this paper we study the influence of the transition metal atom (Mn, Fe, Co and Ni) doping in Bi$_{2}$Se$_{3}$.
Strong influence of the layered structure of the base material into electronic properties is observed.
The quintuple layer structure of Bi$_{2}$Se$_{3}$, which is responsible for an existence of strong bonding between $p$-like orbitals along the five-atom linear chain, plays a crucial role not only in magneto-electric phenomena but also influences the behavior of the topological state of the system.
The introduction of transition metal into the parent material leads to the modification of such bonds.
The strongest changes of distribution of the charge and spin densities are observed along these bonds.

Concluding, the substitution of the the transition metal has a bigger effect on the nearest Se atoms within quintuple layer than on Bi atoms.
Doping by Fe, Co and Ni, shifts the Fermi level to the conduction band.
Moreover, in the case of Co and Ni doping, energy levels located at the Fermi level are mostly associated with the substituted atoms.
This shows that in some cases it is possible to realize  the spin-polarized resonant energy level around the Fermi level~\cite{biswas.balatsky.10}. 
In combination with the relatively large topological gap of Bi$_{2}$Se$_{3}$, this opens a new way of the realization of artificial in-topological-gap bound states induced, in particular, by Co and Ni atoms.
Similarly, in some range of doping by the Mn and Fe atoms, the doping can lead to realization of the new topological behavior associated with quantum Hall-like phenomena or spin-polarized protected edge states.
Both aspects are very interesting from the experimental point of view and possible practical implementation of the doped TI in the spintronic devices.
The further studies of this issue are necessary.

\begin{acknowledgments}
This work was supported by the National Science Centre (NCN, Poland) 
under projects: 
OPUS no. 2017/25/B/ST3/02586 (A.P.), 
SONATINA no. 2017/24/C/ST3/00276 (K.J.K.), 
and 
SONATA no. 2016/21/D/ST3/03385 (A.C.). 
A.P. and K.J.K. appreciate also founding in the frame of scholarships of the Minister of Science and Higher Education (Poland) for outstanding young scientists (2019 edition, nos. 818/STYP/14/2019 and 821/STYP/14/2019, respectively).
\end{acknowledgments}

\bibliography{biblio}

\end{document}